# Four Ways to Scale Up:
# Smart, Dumb, Forced, and Fumbled

By Bent Flyvbjerg[1]

*Saïd Business School Working Papers*

January 2021

Comments very welcome

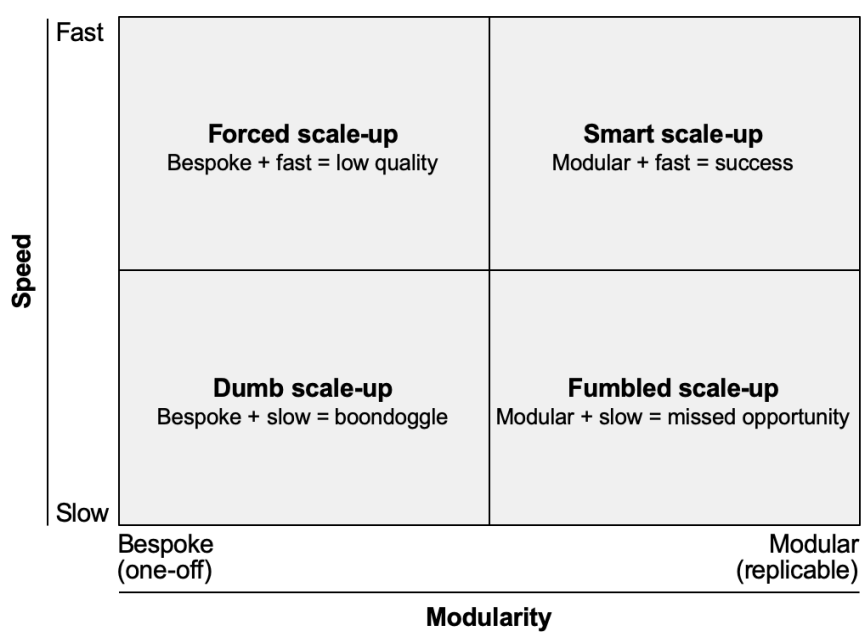

**Full reference**: Flyvbjerg, Bent, 2021, "Four Ways to Scale Up: Smart, Dumb, Forced, and Fumbled," *Saïd Business School Working Papers* (Oxford: University of Oxford).

---

[1] Bent Flyvbjerg is the first BT Professor and inaugural Chair of Major Programme Management at the University of Oxford's Saïd Business School and a Professorial Fellow at St. Anne's College, Oxford.

**Abstract**

Scale-up is the process of growing a venture in size. The paper identifies modularity and speed as keys to successful scale-up. On that basis four types of scale-up are identified: Smart, dumb, forced, and fumbled. Smart scale-up combines modularity and speed. Dumb scale-up is bespoke and slow, and very common. The paper presents examples of each type of scale-up, explaining why they were successful or not. Whether you are a small startup or Elon Musk trying to grow Tesla and SpaceX or Jeff Bezos scaling up Amazon – or you are the US, UK, Chinese, or other government trying to increase power production, expand your infrastructure, or make your health, education, and social services work better – modularity and speed are the answer to effective delivery, or so the paper argues. How well you deal with modularity and speed decides whether your efforts succeed or fail. Most ventures, existing or planned, are neither fully smart nor fully dumb, but have elements of both. Successful organizations work to tip the balance towards smart by (a) introducing elements of smart scale-up into existing ventures and (b) starting new, fully smart-scaled ventures, to make themselves less dumb and ever smarter.

## How to Build 20,000 Schools in the Himalayas

My first megaproject was Nepal's Basic and Primary Education Program (BPEP). This was a billion-dollar venture to improve schooling across Nepal, including the construction of 20,000 new schools and classrooms, during the 1990s and 2000s. I had been hired by Danida – the Danish International Development Agency – to work as program planner on the project and now found myself installed at the Hotel Yak and Yeti in Kathmandu, with responsibility for designing and programming the BPEP.

Because the Himalayas cover most of Nepal, what we economic geographers call "friction of distance" is extreme here. In most places it is difficult to get from point A to point B – including for children to get to school. Roads are absent and many children had daylong walks along difficult mountain trails, often over one or two mountain passes, to get to the nearest school. This meant that many children did not attend school or did so only sporadically.

Nothing is more important, however, for developing a nation, or a family, than getting young children to school, and especially girls. This is because primary education is an effective way to start a virtuous circle of better health, lower birth rates, economic growth, and better living standards. Nepalese villagers were acutely aware of this, as was their government. They wanted schools (Wal 2006: 62). It was therefore a main objective of the BPEP to build the 20,000 schools and classrooms as close as possible to children without access to schools.

On my first night in Kathmandu, the Danish ambassador to Nepal invited me to a private dinner at his home together with Nepal's permanent secretary of education. The permanent secretary and I hit it off, sharing good laughs and our joy of fathering daughters. I now had a direct line to the top of the Ministry of Education, which owned the BPEP and was ultimately responsible for its delivery. This greatly facilitated developing the program and having it approved at all levels of government.

Working with my close collaborator, fellow Dane Hans Lauritz Jørgensen, we set out to detail the program. As an architect, Hans Laurits did the design of individual school buildings and classrooms. As a planner, I did the programming of what had to be built where and when, to deliver the program according to its schedule and objectives. The Nepalese economy is small and fragile. If construction of all 20,000 schools started at the same time, the economy would overheat, a classic error in big construction programs in developing nations. This had to be avoided, pushing the program ahead as swiftly as possible, but not faster than the national economy would allow without being impaired.

After considering different designs for individual school buildings, Hans Lauritz and I settled on three basic prototypes – each type dependent on the gradient of the mountain on which it would be built, with the gradient being a main constraint on design. Turnkey projects would have been an obvious choice, but we decided against them, because studies and site visits showed poor maintenance and lack of local ownership if villagers had not been involved in the construction of their schools. Without proper maintenance, in Nepal's harsh climate schools would deteriorate fast, which would be counter-productive. It was therefore decided that each school would be built with local labor and would involve the local community in decision making, construction, and maintenance.

Finally, we made sure the designs were earthquake proofed, which traditional schools in Nepal were not. If taken into account from the outset, earthquake proofing is inexpensive for this kind of building, so we saw no reason not to do it. This proved a wise decision when, in April 2015, the biggest earthquake since 1934 hit Nepal and killed nearly 9,000 people and injured close to 22,000, with its epicenter in areas with BPEP schools. The news was devastating, but we took solace in the knowledge that children in schools that had been built to our designs would have been well protected, because the buildings would not have collapsed easily.

On behalf of Danida, we negotiated a first draft of the BPEP construction program with JICA, the Japan International Cooperation Agency, which coordinates Japanese governmental development assistance, and which would donate building materials for the schools. After adjustments, the proposal was further negotiated with the World Bank and sufficient funds were raised to begin program delivery. The first 19 districts were covered from 1992 to 1993, adding six more from 1993 to 1994, with a further 15 from 1994 to 1995, and the rest from 1999 to 2004, covering all 75 districts in the country (Chandra 2003: 177).

The BPEP has been, and is, a significant success, according to independent evaluations (Chandra 2003; Jerve, Shimomura, and Hansen 2008; Little 2007; Skar and Cederroth 2005; Wal 2006). A review of development aid to Nepal, commissioned by the Norwegian Foreign Ministry and undertaken by the Nordic Institute of Asian Studies at Copenhagen University, concludes for the first 40 districts of the program:

> "The BPEP has had an enormous impact in a short space of time. It has provided additional classroom space for about 300,000 students and non-formal education for more than 170,000 adults and children. With such achievements it is not surprising that the government is eager to finance a second stage of this project" (Skar and Cederroth 2005: no pagination).

Chandra (2003: 177) evaluated the BPEP management structure and found it has "led to the efficient delivery of project inputs." School enrollment has increased, as intended, and it has increased more for girls than for boys, as also intended, although girls still lag behind in terms of enrollment. Repetition rates (having to repeat a class due to poor performance) have decreased, and again more for girls than for boys (Wal 2006: 63, 77).

Finally, Skar and Cederroth (2005: 49) found that "DANIDA's involvement in the BPEP is regarded as a model by other bilateral donors" and that "donors have flocked" to the BPEP for this reason, helping to raise the necessary funds to make the program a success, including funding from UNICEF, UNDP, the European Union, Norway, and Finland, in addition to the Nepalese government and the funding already mentioned. In total, the program has impacted education for more than a million children and grownups across Nepal.

**The Key to Successful Scale-Up**

To be honest, I did not know much about scale-up when Hans Lauritz Jørgensen and I planned the BPEP construction program for the Nepalese government. Later, when I learned from Danida that the program was showing signs of success, I was delighted but did not think much about it. I thought this was normal, just another day at the office for the megaproject planner. Only when I began to research megaproject performance in my job as a university professor did I realize that success was the exception for such projects and failure the norm. "Over budget, over time, under benefits, over and over again," as I coined the situation in the Iron Law of Megaproject Management, when I later had the data for a full review (Flyvbjerg 2017a: 12).

I then returned to my experience in Nepal and began to ponder what had happened here. Was there a method to our success with the BPEP? Or had we just been lucky? Was it the plentiful and timely funding from donors that did it? Or the unwavering support from the permanent secretary and his ministry? Or the way Danida had managed things?

It took me a while to figure out the answers to these questions. First, I needed experience and research from more projects. Ultimately, I found that the BPEP success was explained by none of the above factors, although they are all important. Instead, I found that the following two features of the BPEP were the keys to its success, more than anything else:

1. The program was designed and delivered in a *modular, replicable* fashion.
2. The program was designed and delivered at *speed*.

First, regarding modularity and replicability, the constraints of the Nepalese geography, the scarcity of labor and materials, and the vast scale of the program pushed us to be frugal, working with just a few standard designs that could be easily built and replicated over and over, as LEGOs in the Himalayas.[2] In essence, officials and villagers would choose between two or three standard designs of schools – which would be the basic building blocks of the BPEP – literally determined by the slope of the mountain on the site where the school would be built.

You would then build the chosen design, following relevant instructions, and populate the building with teachers, teaching materials, and children. You would repeat this until a district was covered. Then you would do the same in the next district, and the next, until all 75 districts in the country had been included. Replicability applied not only at the level of the building modules (the schools and classrooms), but also at the district level, where experience from one district would be replicated in the next, with the district now being the module to repeat over and over.

Research shows that replicability is crucial to effective *learning* (Baldwin and Clark 2000). Replicability creates a feedback loop where you can use the experience from delivering one module to improve the delivery of the next, repeatedly, ensuring that the quality of delivery constantly improves as you go along, due to learning-by-doing. Replicability is also conducive to *experimentation*. Instead of going full scale immediately, you experiment with a few modules and use your experience from the experiments to improve the next modules, and you repeat this until you master delivery, which is when you go full scale.

The ability to experiment and learn – something humans are inherently good at, probably because it enhances our chances of survival in evolutionary terms – is the most basic explanation why a venture that is based on modular replicability is more likely to succeed than a venture that depends on a one-off, bespoke construct that can only be delivered in one go – something humans are inherently bad at, having difficulty getting things right the first time. Experimentation and learning are main reasons the BPEP succeeded instead of failing like the majority of megaprojects.

---

[2] Modularity in construction often conjures up images of unattractive, low-quality buildings, like 1960's Soviet, US, and UK social housing. Here we maintain that modular construction can and should be aesthetically pleasing and of high quality, and that there is no good reason for this to not be the case, but quite the opposite. Aesthetics and quality were not compromised by modularity for the 20,000 schools in Nepal.

Second, regarding speed, it was clear that Nepal immediately needed more schools, which the permanent secretary had hammered into me when we first met at the ambassador's house. Speed was therefore of the essence. Neither design, negotiations, decision making, nor delivery could be allowed to drag on for years, as is common for large programs. It took us only a few weeks to develop the first draft of the construction program, not including negotiations for funding. The other parts of the BPEP were similarly accelerated and developed in parallel. Raising funds and final decisions took a few months and construction began immediately after this, with 19 districts covered in just a year, which is exceptionally fast for this type of project.

It should be emphasized that despite the accelerated pace, the BPEP was not fast-tracked. With fast-tracking, construction starts before designs and plans are completed, which is common when speedy delivery is a must. However, fast-tracking is notoriously a high-risk strategy, because the chances of making wrong decisions multiply without a firm design (Williams and Samset 2009, Tighe 1991). For the BPEP such risk was not an option, so designs and plans were completed before construction began and proved sufficiently robust to keep the program on track throughout.

Two things, in particular, contributed to the speedy delivery: (a) the simple, modular designs for schools, which were easy and quick to build, making replicability a facilitator of speed, and (b) the fact that a number of districts were completed within just a year, after which finishing the program became a matter of repeating the experience from these districts in other districts, over and over, until the whole country had been covered, again making replicability pave the way for speed. In fact, replicability at the level of both individual buildings and individual districts worked so well that the whole BPEP was delivered years ahead of schedule, a rare feat for programs of this size (Flyvbjerg 2017a).

The success of Nepal's BPEP was beginner's luck in the sense that I was fortunate that my first megaproject lent itself to modular replication and accelerated speed. We did not realize at the time just *how* lucky we were and how rare success is in megaproject management, because nobody knew at the time, for lack of systematic evidence. However, the BPEP success was *not* beginner's luck in the sense that we quickly realized that modularity, replicability, and speed would be key to success, and we proactively and very deliberately designed the program around these concepts, to great effect.

Today, I have learned – through my experience from working on dozens more megaprojects and from my research at Aalborg, Delft, and Oxford on the performance of thousands of projects – that modularity, replicability, and speed are *the* key determinants of success, not just in programs like Nepal's BPEP, but in any type of megaproject or other large-scale venture.

Whether you're a small startup or Elon Musk trying to grow Tesla and SpaceX, or Jeff Bezos scaling up Amazon, or Larry Page and Sergei Brin trying to cover the planet with Internet and Google servers – or you're the US, UK, Chinese, or other government trying to increase power production, expand your infrastructure, or make your health, education, and social services work better – modularity, replicability, and speed is the answer to effective delivery. How well you deal with these three issues will decide whether your efforts succeed or fail.

Today, the first thing I do when assessing whether a prospective megaproject – or other outsized venture – is likely to succeed, is to assess the degree to which it lends itself to modularization, replication, and speed. *If a project can be delivered fast in a replicable, modular manner, it is likely to do well. If it cannot, it is likely to be troubled or fail.*

**Four Ways to Scale**

Figure 1 illustrates the importance of modularity and speed with the help of a simple two-by-two matrix. On the horizontal axis, the figure shows modularity, with projects grouped as either modular and replicable or bespoke and one-off. On the vertical axis, the figure shows speed, with projects grouped as fast or slow. This results in four main types of scale-up, of which smart scale-up and dumb scale-up are the most important.

We define *scalability* as the capacity to change something in size or scale, for instance, the potential for an entity to be able to handle a growing – or diminishing – workload. Contrary to common understanding, being big is not the same as being scalable. Scalability is the capacity to *change* in size, including the capacity to grow big at speed, but scalability is not identical to size in itself. True scalability resides in the smart quadrant of Figure 1 and dumb scale-up is not scalable, it's just big.

For instance, the Channel rail tunnel between France and the UK has a fixed capacity that is not scalable, and because the tunnel has only had demand at just over half of this capacity the investment proved financially and economically nonviable. Similarly, consider a large hydroelectric dam. It cannot scale beyond its capacity and is therefore not scalable. Now contrast this with a wind farm where you can add, or remove, turbines to adjust capacity as needed, up or down. The wind farm is scalable. And if you need more capacity than one wind farm can provide, you can add another wind-farm in the same manner, so that windfarms are scalable not only at the level of turbines but also at the level of farms, exactly like the schools and the districts were both scalable in Nepal. The same holds for server farms, solar farms, energy storage and transmission, and many other phenomena. Wind is just an example. We call the ability to scale at any scale "*scale-free scalability*." It is a core feature of truly scalable systems.

In what follows we account for each of the four types of scale-up, one by one.

*Figure 1: Four ways to scale up, based on modularity and speed*

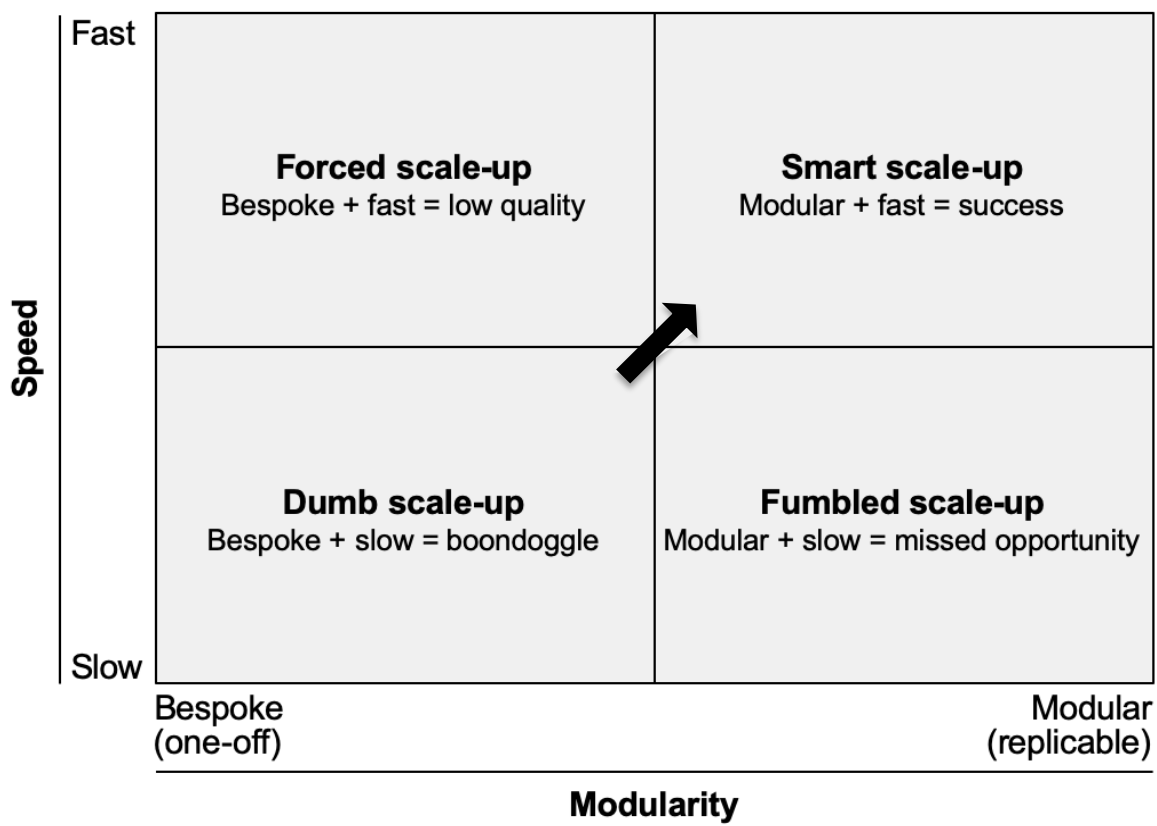

## Smart Scale-Up

Smart scale-up is placed in the upper right-hand quadrant of Figure 1. This is the type of scale-up we lucked upon in Nepal with the BPEP. Smart scale-up entails developing something big based on the principles of modularity, replicability, and speed, as described above.

 

To take another example of smart scale-up, as different from Nepal's BPEP as can possibly be, consider the Tesla Gigafactory 1, also known as Giga Nevada. This is a five-billion-dollar high-tech lithium-ion battery factory under construction in Nevada, which is intended to help make electric vehicles and home-power systems more affordable by producing batteries at an unprecedented scale and thereby bringing down their cost. Gigafactory 1 will be the biggest building in the world by footprint, at more than half a million square meters, or 107 football fields. The building is modular by design. The basic module is called a "block" and is a part of the factory that can be operated as soon as it is completed, while more blocks are being built that can also be operated when finished, and so on until the whole factory has been completed with production ramped up block by block. The block is the LEGO of the Gigafactory, just like the individual school was the LEGO of the BPEP.

Construction of Gigafactory 1 started in late 2014. By the third quarter of 2015, a first portion of the factory had been finished and the Tesla Powerwall (a home energy storage system) began production here (Whaley 2016). By July 2016, the grand opening event for the factory was held, with three of 21 blocks completed, amounting to approximately 14 percent of the final total size (Weintraub 2016). Mass production of battery cells began in January 2017 (Randall 2017). The accelerated pace at Gigafactory 1 is different from what you normally see in construction, where a project of this cost and size would typically take five to seven years before operations begin.

Tesla reaped two substantial advantages from its emphasis on speed. First, the company exposed itself to lower risks of cost overrun, because the likelihood of cost overrun correlates positively with length of schedule. That's smart. Second, the company got to their revenue stream and started serving customers years earlier than would have been the case following the conventional approach. That's even smarter, which is why we call Tesla's approach "smart scale-up." Both of these advantages are crucial to a fast-growing company that needs cash in order to grow and cannot afford to have its funds tied up in slow-moving, risky construction projects, which would be dumb.

The CEO of one of the world's oldest and largest construction companies told me an interesting story about his experience with Gigafactory 1. When he heard of Tesla's plans to build the factory, he called Tesla CEO Elon Musk to offer his services, CEO to CEO. "We know a thing or two about the construction of large projects and thought we could help," the CEO told me. Musk did not answer and did not return the calls, which irked the CEO. He later learned that Musk wanted Tesla to reinvent the construction process for Gigafactory 1 from basics – like he had previously reinvented the produc-tion of cars and rockets – and that he did not want to risk being influenced by conventional thinking by involving a traditional construction company.

The CEO was understandably offended by not having his calls answered. But even more, he was shaken. Was this a first glimpse of the future of construction, disrupted by high-tech entrepreneurs like Musk, the CEO had to wonder? And he was right to be concerned. As the least productive (and most corrupt) sector of the economy, construction is long overdue for disruption. Musk and other tech entrepreneurs know this and it is only a matter of time before they, or someone else, do something about it. If you work in construction you are well advised to pay close attention to what is going on at Tesla's Gigafactories and in Silicon Valley. You should also consider running your business based on the principles laid out in this paper, so you may disrupt yourself before others disrupt you.

At first glance, nothing could seem more different than Tesla building its high-tech Gigafactory in Nevada aimed at global disruption of the energy and transportation industries and Nepal's Ministry of Education building low-tech primary schools in remote parts of the Himalayas aimed at getting village children, and especially girls, to attend school. If you look closer, however, there are crucial similarities, and the similarities are what explain the success of each venture.

First, both Tesla and the Ministry of Education took a modular approach to the design and construction of their projects, Tesla with its "blocks" of factory space, the Ministry with its standard school designs. Each module – whether a block or a school – was immediately operational upon completion, minimizing lead time from decision to build to actually delivering products and services.

Second, Tesla and the Ministry both emphasized replicability, Tesla in building one factory block after another and the Ministry in building one school after another. In both cases this made effective learning possible so that later modules could benefit from the experiments, successes, and failures experienced for previous modules and thus make delivery more effective than it would otherwise have been. "Fail fast – and learn from it," was just as pertinent for Nepal's low-tech schools as for Nevada's high-tech Gigafactory.

Speed was therefore of the essence for both Tesla and the Ministry. With its fast and frugal school designs, the BPEP had happened upon what lean startups would later call a Minimum Viable Product (MVP). This is a product with just enough features to start initial rollout with the purpose of accelerating learning about the product for further development and rollout. For Tesla, its blocks of factory space at Gigafactory 1 similarly defined a minimum viable production facility that could begin opera-tions as soon as it was completed and would deliver learning for the following blocks. As a consequence of this approach, both Tesla and the Ministry began operations and delivered their first products and services within just one year of deciding to go ahead with their respective projects. Such speed is highly unusual for ventures of this type and size. For Tesla, in addition to facilitating learning, the speedy approach also meant quick access to much-needed revenues for further growth to achieve its ambitious aims. For Nepal's Ministry of Education it meant that more children were attending school with almost immediate effect, which was the very reason for the BPEP.

---

**Wind: Getting Speed Right**

When the British prime minister inaugurated the London Array in July 2013, it was the largest offshore wind farm in the world, costing USD 3.0 billion in 2012 prices (Sovacool et al. 2016). Located in the Thames Estuary, with 175 turbines and a capacity of 630 MW, the farm delivered enough power to supply 500,000 British households with clean electricity. Construction began in March 2011, electricity production in October 2012, and all turbines were confirmed fully operational by April 2013, just two years and a month after construction started (Lacal-Arentegui et al. 2018). Delivering multi-billion-dollar infrastructures at such speed is highly unusual, and onshore wind is even faster. Conventional power plants, nuclear plants, hydropower dams, and even standard large-scale infrastructures such as rail lines, motorways, tunnels, bridges, etc. routinely take years longer. This despite the fact that offshore construction faces some of the harshest weather and sea conditions that exist. By today's standards, delivery of the London Array was not even particularly fast. Since 2011, the speed of installing offshore megawatts has accelerated on two fronts. First, the speed of installing individual turbines has increased, with some companies now claiming one-day installations. Second, and more importantly, the average size of installed turbines has grown 3 to 4 times since 2011. As a result of the two trends, installation days per MW have dropped dramatically and the holy grail of being able to deliver a full offshore wind farm in just one year has been achieved (Lacal-Arentegui et al. 2018: 4, 11). The London Array, and contemporaneous farms, took minimum two years to build, with costly and complicating intermissions during the winter. The 2018 Walney Wind Farm extension, located off the West Coast of England with 87 turbines and a capacity of 659 MW, is an example of a farm built in less than year, which is becoming the new industry standard, bringing down cost and risk. Such speed is rare for multi-billion-dollar megaprojects. Other project types have much to learn from wind and are well advised to study this industry carefully.

---

## Why Speed Matters

There is a simple, yet profound, reason why speed is critical to smart scale-up and to success. We call it the First Law of Forecasting:

> First Law of Forecasting: *You have relative certainty for the first year of a forecast, and you can forget about knowing much about anything beyond three to five years*.

The law explains why it was such a brilliant move on the part of Tesla and Nepal's Ministry of Education to insist on going live with delivering their product within the first year and to deliver as much as possible within the first few years. At the most fundamental level, the longer it takes to achieve an outcome, the more variance this outcome will have, and variance translates directly into uncertainty and risk. Keeping things short is therefore a basic and direct way to reduce risk, and reducing risk is key to success for large projects

The importance of time to forecasting accuracy has been documented by Tetlock (2005) and Tetlock and Gardner (2015). Through decades of research Tetlock found that the time horizon within which humans can forecast events like political elections, GDP growth, macroeconomic policies, business cycles, developments on the internet, geo-political developments, ethnic conflict, state collapse, wars, peace, etc. is relatively short. For the first year, forecasting accuracy is fairly high, but after this accuracy declines rapidly, to disappear into the mists of randomness at a forecasting horizon of three to five years and beyond. This holds for even the best forecasters, called "superforecasters." Tetlock summarizes his finding like this:

> "It was easiest to beat chance on the shortest-range questions that only required looking one year out, and accuracy fell off the further out our experts tried to forecast – approaching the dart-throwing-chimpanzee level [i.e., randomness] three to five years out. That was an important finding. It tells us something about the limits of expertise in a complex world – and the limits on what it might be possible for even superforecasters to achieve" (Tetlock and Gardner 2015: 5).

It should be mentioned that Tetlock's conclusion might be optimistic, that is, the time horizons within which most real-life forecasts are accurate might be shorter than those he identify. This is because, first, Tetlock's results are based on particularly skilled forecasters, as he mentions. Second, and perhaps more importantly, the forecasts Tetlock studies are simplified forecasts, often with a binary answer, of the type used in so-called "forecasting tournaments," which are key to Tetlock's research. Here participants compete against each other with their performance measured by who can most accurately forecast questions like: "In the next year, will any country withdraw from the eurozone?" or "Will North Korea detonate a nuclear device before the end of this year?" (Tetlock and Gardner 2015: 2). Many real-life forecasts do not have binary answers, however, but instead ones that vary over a wide range of possible outcomes, for instance: "How many people are likely to die from covid-19 over the next year?" or "How much is the California high-speed rail system likely to cost?" The first type of question is easier to forecast than the second, but the second is probably more common, outside the simplified world of forecasting tournaments.

It should also be mentioned that the First Law of Forecasting is an overgeneralization, because different time horizons for accuracy will apply to different types of forecasts. A weather forecast, for example, has only short-term accuracy, up to ten days. Beyond that, human forecasters are no better than Tetlock's dart-throwing chimpanzee. A forecast of life expectancy, in contrast, can be made with much higher accuracy and for significantly longer forecasting horizons, covering several decades.

Finally, it should be mentioned that the First Law applies only to phenomena that are actually forecastable, which some phenomena are not. Phenomena with regression to the mean (finite mean and variance) are generally forecastable while phenomena with regression to the tail (infinite, i.e., nonexistent, mean and variance) are difficult or impossible to forecast reliably, and it is typically a fool's errand to try (Flyvbjerg 2020). Other approaches than conventional forecasting must be applied for such phenomena, for instance extreme value theory and black swan management (Taleb 2012: 306-307; Flyvbjerg and Budzier 2011; Haan and Ferreira 2006). This leads to what we will call the Second Law of Forecasting:

> Second Law of Forecasting: *You should only attempt to forecast that which is actually forecastable, and never pretend something is forecastable that is not.*

The basis for the second law is well documented by Mandelbrot (1997), Mandelbrot and Hudson (2008), and Taleb (2019, 2020).

Despite the limitations of the First Law of Forecasting, this law can be immensely helpful for getting scale-up right in domains where the law applies. In such domains, what the law tells leaders in business, government, and NGOs who are trying to scale up a product or a service is that to maximize their chances of success they must:

a) Have a Minimum Viable Product and have started to deliver this within the first year of deciding to go ahead with it; and
b) Have completed delivery or, at a minimum, a substantial part of this (and, for businesses, established themselves as a market leader) within three to five years. Anything beyond this time horizon will be more or less random, that is, luck, and, therefore, much less likely to succeed.

Leaders who practice smart scale-up and have a record of delivering successfully at scale religiously observe the first and second laws of forecasting. Leaders who ignore these laws – and many do, as we will see below – do so at their own peril and often come to grief.

Speedy scale-up is all the rage in tech companies around the world, with Silicon Valley as the epicenter. This is because competition in tech is characterized by winner-takes-all, so speed is of the essence for capturing market quickly to outmaneuver the competition and become a dominant player. Former Alphabet chairman and CEO Eric Schmidt and former Google senior vice president of products Jonathan Rosenberg explain:

> "Create a product, ship it, see how it does, design and implement improvements, and push it back out. *Ship and iterate*. The companies that are the fastest at this process will win" (Schmidt and Rosenberg 2015: 234, italics in the original).

LinkedIn co-founder Reid Hoffman calls this process "blitzscaling" and argues that *scale-ups*, and *not* start-ups, are what distinguish Silicon Valley compared with other tech ecologies (Hoffman and Yeh 2018). It is the better conditions for blitzscaling in Silicon Valley – which emphasize and support speed over anything else for achieving success in growing tech companies – that make the Valley hard to compete with for other tech clusters, according to Hoffman.

Here we argue that speed is only half the story, however, albeit an important half. Leaders also need to know how to most effectively attain and manage speed, or speed will risk derailing their venture – for instance if they think that fast-tracking is an effective way to attain speed, which is common but mistaken. This is why we emphasize *modular replicability* as the other half of the story, in addition to speed. Modularity and replicability are the most effective means to achieve and manage speed in a stable manner, that is, without getting into a situation where higher speed leads to higher risk, as with fast-tracking, but quite the opposite. We maintain that modularity, replicability, and speed are all preconditions of success. It will not do to focus on just one of them, as blitzscaling does. If you don't know what your LEGO is and how to build with it, you cannot blitzscale. As a leader, you need to have a good answer to the question: "*What's my LEGO?*"

Even beyond the three to five years for which forecasting works, according to the First Law of Forecasting, if you have spent those years replicating a modular product at speed, you are much more likely to succeed beyond this period than if you had spent those years planning, designing, and devel-oping your bespoke product, getting ready for a big-bang launch, which is the common approach for large, capital-intensive ventures, but which is dumb.

In sum, the secret to successful scale-up – doing something big successfully – is fast, modular replication. Conversely, the road to failure – which is taken all too often – is deciding to do a big, bespoke venture, and do it slowly. We call the latter dumb scale-up, which we consider next.

---

**Dove Satellites: A Case of Smart Scale-Up**

Will Marshall was a young engineer working at NASA's Jet Propulsion Laboratory (JPL) in California building big, bespoke spacecraft in the tradition of last century's moon missions. Eventually, he got tired of the slowness and waste of Big Space and decided to do things differently. Along with two other NASA alumni, he started his own company and built a satellite called Dove in his garage in Cupertino. During the 2010s, the company, Planet Labs, launched several hundred satellites, forming the largest constellation ever put into orbit, providing up-to-date information for climate monitoring, farming, disaster response, and urban planning. At a weight of ten pounds, a build-time of a few months, and a cost under a million dollars, including launch and operations, a Dove satellite is radically smaller, faster, and cheaper to build than anything at NASA, but they are equally well engineered and more agile. Each satellite is made up of three so-called CubeSat modules, which are themselves made up of multiples of 10x10x10 cm modules – Marshall's LEGOs – using standard commercial off-the-shelf components for their electronics and structure, like those mass-produced for cell phones and recreational drones, keeping cost and delivery times low. Marshall remembers how he lost twenty-six Dove satellites in 2014. They were sitting on a big rocket that exploded on the launch-pad. The loss hardly affected his business, since he had had nine successful launches and only one failure. The lost satellites were quickly replaced and the replacements put in orbit. Marshall's modular approach means that every mission is cheap enough to fail and fast enough to replicate in case of failure, with lessons from the failure being immediately useful for the next iteration. In contrast, missions in the NASA Big Space tradition are too big to fail and too slow to replicate when they do. NASA typically takes a decade to plan and a decade to build its bespoke designs. The longer the time the higher the risk of ultimate failure, with little to learn from failing, because technology has moved on and "fail slow" mostly does not work. As a result, Big Space is ready for disruption, as rightly seen by Marshall. Bigger players, like Elon Musk's SpaceX and Jeff Bezos's Blue Origin, share the vision and promise to disrupt Big Space even more, based on the same principles as those used by Marshall: dramatically lowering costs and delivery times based on the use of industrially manufactured standard building blocks (Tepper 2015, Dyson 2016).

---

## Dumb Scale-Up

Outside tech companies, smart scale-up is rare. In conventional business and government, dumb scale-up is common, with devastating consequences for investors' and taxpayers' money, and for the wealth of nations (Ansar et al. 2014, Detter and Fölster 2015, Kanter 2015).

Dumb scale-up – placed in the lower left-hand quadrant of Figure 1 – is the opposite of smart scale-up. Here you slowly build something big and bespoke, for instance a nuclear power plant, a large hydro-

                                                                                                                                                                                                                                                                                     

electric dam, a big-bang enterprise resource planning (ERP) system, a chemical processing plant, a new mine, a national health or pension IT system, a big defense system, a space shuttle, an aircraft, or a new airport. Dumb-scaled ventures are often binary, in the sense that they are either completely on or completely off, with no middle ground. For instance, a dam or a nuclear reactor cannot be completed in incremental fashion, like the Gigafactory or the schools in Nepal described previously. Even a 95 percent completed dam or nuclear reactor is of no use; it must be 100 percent completed before it can deliver its benefits. That's a big drawback for anyone who needs to impact the world now, or get to the cash flow fast to grow their business.

Dumb scale-up strongly correlates with high costs and large cost overruns, long schedules and schedule delays, and sizable benefit shortfalls (Flyvbjerg 2014, 2016). The combination of cost overrun and benefit shortfall makes for bad business, needless to say, which is why this approach is called "dumb": it wastes money, talent, and other resources. The result is your classical boondoggle, like California's high-speed rail system, Japan's Monju nuclear power plant, the New South China Mall, the F-35 Joint Strike Fighter program, Kmarts enterprise resource planning system, the UK National Health Service IT system, the International Space Station, Pakistan's Tarbela dam, or Berlin's Brandenburg airport. In the terminology of Taleb (2012), dumb-scaled ventures are *fragile* whereas smart-scaled ones are robust, or even antifragile (Ansar et al. 2017).

Consider Japan's Monju nuclear power plant in more detail. Monju is a prototype fast-breeder reactor, the first of its kind for commercial use. It was intended to become the cornerstone of a high-priority national program to reuse and eventually produce nuclear fuel in a country with few energy sources of its own (Tabuchi 2011, Phys.org 2014). The reactor was named after the Buddhist deity symbolizing wisdom. The decision to build Monju would prove anything but. Construction was approved by the High Court in 1983 and began in 1986. With the long gestation period typical of nuclear power, inauguration took place 12 years later, in 1995.

A few months after inauguration, in December 1995, a major fire, possibly caused by defective welding, shut down the facility (World Nuclear News 2016). An attempt by the operator to cover up the accident – involving the editing of videotapes, falsifying reports, and issuing a gag order on employees – developed into a political scandal with massive public outrage. A five-year delay ensued, but in November 2000, finally the Japan Atomic Energy Agency announced they would restart Monju. The announcement was met with protests and court battles, resulting in further delays.

Not until May 2005 did Japan's Supreme Court give permission to reopen Monju. Restart was scheduled for May 2008, but was postponed to October, which was then postponed to February 2009, which was again delayed when holes were discovered in the reactor's auxiliary building. Finally, in August 2009 a restart for February 2010 was announced. In March 2010 test runs began that were to continue until 2013, at which time the reactor was planned to begin full operations, feeding power into the electric grid – 30 years after approval. However, on August 26, 2010 a 3.3 ton machine for refueling fell into the reactor vessel. October 2010 saw a retrieval attempt, but the machine had become misshaped by its fall and no longer fit through the lid of the reactor, so the attempt failed. In June 2011, after months of engineering work, the machine was finally retrieved (Mainichi Daily News 2011; Kubota 2011).

In May 2013, Monju was ordered to suspend its preparations for restarting the reactor, after maintenance flaws were discovered on some 14,000 components at the plant, including safety-critical equipment. Additional uninspected equipment was identified in 2014, together with more than one hundred improper corrections to inspection records, leading to suspicions that the records were being tampered with. In 2015, it was further found that mandatory periodical degradation measurements of the thickness of the cooling pipes had not been carried out for the past eight years. The Nuclear Regulation Authority declared the operator of Monju not qualified to operate the reactor. The government found that it would cost an additional 6 billion dollars to restart Monju and operate it for ten years, on top of the 12 billion dollars already spent (Japan Times 2014, 2015; Green 2016; World Nuclear News 2016).

After more than 20 years of incident after incident and delay upon delay, with no electricity or revenues to show for the troubles, the government was finally losing patience. Moreover, the Fukushima nuclear disaster in March 2011 had increased public resistance to nuclear power and had made the government reconsider its nuclear policy. Remember the laws of forecasting above? Planning 20 to 30 years ahead is a fool's game. Time is like a window. The bigger you make it, the more risks can fly through it. Including big, fat black swans like Fukushima.

In December 2016, the Japanese government decided that Monju – once considered the cornerstone of Japan's nuclear policy – would have no further role in fulfilling this policy. It was decided to close and decommission the Monju reactor (Japan Times 2016, BBC 2016).

Logically, this decision would have been followed by a decision to scrap the partially-completed Rokkasho reprocessing plant, designed to provide plutonium fuel to Monju and other planned fast-breeder reactors, fuel that would now not be needed. Interestingly, this did not happen, perhaps indicating that Japan was not quite ready to follow the lead of the US, the UK, and Germany, who had all cancelled their prototype breeder reactor programs. Like Monju, operations at Rokkasho were delayed over and over, in total by 25 years. At the time of writing, the estimated opening year was 2022, after deadlines in 2013, 2014, 2015, and 2018 had been missed. The current estimated cost is 22 billion dollars, three times the original estimate. Rokkasho seems on track to become a giant white elephant, with no need for its product (Yamaguchi 2015, Green 2016, Japan Times 2016, World Nuclear News 2017).

In sum, after 34 years of efforts and 12 billion dollars in expenditures – not including the 22 billion spent at the Rokkasho processing plant – Monju has nothing to show for it. No electricity for Japanese consumers, no revenues for business and government, and no new fuel for Japan's other nuclear reactors. Monju is said to have generated electricity for all of one hour during its 22-year lifetime (Tabuchi 2011). That's dumb. And the story does not end here. Decommissioning of Monju is scheduled to take another 30 years, until 2047, at a cost of a further 3.4 billion dollars. If previous experience is anything to go by these numbers are optimistic, with further delays and cost overruns a near-certainty (World Nuclear News 2016). At a minimum, Monju will end up a 60-year, 15-billion dollar venture with zero or negative benefits. Wisdom, anyone? No, truly dumb.

You will be excused if you find it difficult to believe the story of Monju, or if you think it is just an extreme case of bad luck. After all, the story sounds too dumb to be true, or at least too dumb to be common. But the above are the bare facts of what happened, not even touching upon the real drama of the case, which is more extreme still, including the suicide of a top manager at the plant on the day it was announced how much it would cost to recover the refueling machine from the reactor; or the court cases and public protests, which became increasingly vile after Fukushima; or the fact that Monju is built on an active geological fault; or the fear that Monju's troubles would result in radioactive material ending up in the wrong hands and being used by rogue states and terrorists for nuclear weapons. The real tragedy of Monju is that the story of delay upon delay, billions of dollars spent, and little or no benefits to show for it – all caused by dumb scale-up – is all too common (Altshuler and Luberoff 2003, Flyvbjerg 2017b, Hall 1980, Morris 2013).

## Negative Learning

To better understand what went wrong at Monju, let us contrast dumb with smart scale-up.

For Monju, there were no LEGOs for simple assembly into the overall project. Nothing was modular and standardized, everything was bespoke, designed and produced for this and only this one-off prototype. There is nothing at Monju that compares to the replication of the 21 production modules at Gigafactory 1, described above, or the modularity and reproducibility of the 20,000 schools and classrooms in Nepal. At the Gigafactory and in Nepal, those involved learned from the delivery of one module to the next and thus got better and better and faster and faster at what they were doing. In contrast, at Monju, everything was done just once, with great difficulty because it involved extreme complexity and had never been done before. In many instances, the difficulties at Monju were so severe that instead of getting faster as delivery progressed, things got slower and slower, resulting in numerous delays, finally grinding to a halt without ever achieving the planned outcomes.

Instead of the positive and repetitive learning-by-doing that is at the heart of smart scale-up, at Monju learning was adverse in the sense that the more experience that was achieved by those involved, the more they realized that the project would be more difficult, take longer, and cost more than anticipated. The more they learned, the more challenged they became, until they could no longer overcome the challenges. This phenomenon is called *negative learning* and it has been found to apply to nuclear programs not only in Japan, but also the United States, Canada, France, Western Germany, and India (Grubler 2010: 5174, 5186; Escobar-Rangel and Lévêque 2015; Lovering et al. 2016).[3] In the present analysis, negative learning is caused by dumb scale-up, that is, choosing to scale something that does not easily scale, like big nuclear power. *Positive learning* – learning to do things faster, better, and cheaper by doing the same thing over and over, through replicated modules – is facilitated by smart scale-up and is what makes ventures succeed that are based on smart scale-up, like Tesla's Gigafactory and the schools in Nepal.

It is unfortunately not difficult to find more examples like Monju, past or present. Consider the four nuclear reactors recently under construction in the US, the Virgil 2 and 3 reactors at the V. C. Summer nuclear station in South Carolina and the Vogtle 3 and 4 reactors at the A. W. Vogtle plant in Georgia. The reactors, which were estimated by Morgan Stanley to cost 41 billion dollars, are the first new developments of nuclear power in the US since the Three Mile Island accident in 1979 (Hals et al. 2017). They were planned to lead a national renaissance of nuclear power. However, the four reactors suffered all the symptoms of dumb scale-up described above: years behind schedule, billions of dollars over budget, and diminishing benefits rapidly retreating into the future, with devastating consequences for the companies involved, consumers, the nuclear industry, and society as a whole.

---

[3] South Korea is a notable exception from negative learning among countries for which data are available (Lovering et al. 2016: 378). China is not included for want of valid and reliable data.

Ironically, the four US reactors were planned to incorporate elements of smart scale-up through a new design developed by Westinghouse, which was the US nuclear engineering subsidiary of Japanese Toshiba. The new design was less bespoke and more simple than conventional reactor designs, with sections manufactured at factories off site, needing only assembly on site, resulting in a more speedy delivery. Or that was the plan.

In reality, negative learning quickly set in and caught up with the good intentions. Construction of Virgil 2 and Vogtle 3 began in March 2013. Just three months later, in June 2013, a delay of 14 months was announced for Vogtle 3. The year after, Virgil 2 announced a delay of a year and a cost overrun of 1.2 billion dollars, mainly caused by fabrication delays. At Vogtle, contractors installed 1,200 tons of steel reinforcing bars in violation of the approved design, which prompted a seven-and-a half month delay to get a license amendment (Cardwell 2017). Further schedule delays and cost overruns accumulated as it became clear that the new design was all-too similar to previous designs in the sense that it, too, had negative learning and proved significantly more difficult to deliver than anticipated. After a decades-long hiatus in nuclear construction, American companies lacked expertise and equipment. They were simply not up to the job. Some of the work had to be shipped abroad, again causing overruns.

By early 2017, delays and cost overruns on the four US reactors were so out of control that they threatened the viability of not only the reactors, but of the corporations involved. Westinghouse, founded in 1886 to bring electricity to the masses, was forced to file for bankruptcy. Toshiba, the parent company, lost more than half its value and was pushed to the brink of financial ruin. Its chairman, formerly chair at Westinghouse, stepped down and the company decided to withdraw from all overseas nuclear development, previously seen by Toshiba as a growth business with plans to build 45 new reactors around the world (Cardwell 2017, BBC 2017). Given the scale of the collapse, the thousands of jobs at stake, and an 8.3 billion dollar US government loan guarantee provided to help finance the reactors, the bankruptcy threatened to entangle the US and Japanese governments (Hals et al. 2017). *The Wall Street Journal* ran a front-page story with the CEO of Southern Co., owner of the two Vogtle reactors, desperately calling for the help of the US President and the Japanese Prime Minister to save the reactors, stating that "The commitments are not just financial and operational, but there are moral commitments as well" (Gold and Negishi 2017).

In July 2017, South Carolina Electric and Gas reviewed the costs of Virgil 2 and 3, after which they decided to stop construction and abandon the two half-finished reactors, to avoid throwing good billions after bad (Cunningham and Polson 2017). Construction on the Vogtle 3 and 4 reactors continued, with further cost overruns and delays, and with planned in-service dates for November 2021 and November 2022, respectively, at an estimated total cost of 25 billion dollars – and little prospect of ever becoming profitable (Georgia Power 2018, Walton 2018).

The Virgil and Vogtle debacles have dashed the hopes for those who saw the four reactors, with their innovative design, as heralding a new age for nuclear power in the US. If anything, Westinghouse's filing for bankruptcy and Toshiba's pulling back from nuclear construction have set back the nuclear industry by decades, once again.

The situation is similar in Western Europe where the only two reactors under full construction – at Olkiluoto in Finland and Flamanville in France – have even larger delays and cost overruns than their American counterparts. At Olkiluoto, the delay is eight to ten years with a cost overrun of 180 percent; at Flamanville, it is six years and 220 percent, so far (Landauro 2012, Milne 2014, World Nuclear News 2014, EDF 2015, The Economist 2016b). Here, too, there is little certainty about the final outcome in terms of opening date and cost, and default remains a real option, like in the USA. In 2009, Petteri Tiippana, the director of Finland's Radiation and Nuclear Safety Authority, explained to the BBC that the design of the reactor at Olkiluoto was not safe and that thousands of mistakes had been made during its construction. Like in the US, builders were not up to meeting the exacting standards required on nuclear construction sites (Jones, Meirion, 2009). In the UK, the Hinkley Point C nuclear power station was in the early stages of construction at the time of writing, with significant delays and cost overruns incurred even before construction had begun.

Perhaps the problems with nuclear should come as no surprise. After all, General Electric, a pioneer in the field, previously scaled back its nuclear operations, citing doubt about their economic viability. Siemens AG, a German conglomerate, decided to abandon the industry. Areva SA, a French builder, was mired in financial and safety problems. The market value of China General Nuclear Power Group had fallen, and Russia's Atomenergoprom had junk-bond status (Gold and Negishi 2017). What we may be witnessing here is the slow demise of nuclear power as private business. Nuclear may be inherently just too difficult to become profitable on market terms, not least because of the high safety standards that keep escalating with each new major accident, from Three Mile Island over Chernobyl to Fukushima. The escalated standards require ever-more cumbersome, tailor-made, expensive solutions to solve new problems as they arise.

So far this has made smart scale-up of nuclear power impossible, leaving dumb scale-up – bespoke and slow – as the fallback position. "Nuclear safety always undermines nuclear economics. Inherently, it's a technology whose time never comes," as pithily observed by Dr. Mark Cooper of Vermont Law School (Cardwell 2017). The high safety standards demand perfection, but humans are not perfect. Humans learn by tinkering, yet tinkering is too dangerous for nuclear. Nuclear therefore falls short over and over, caught in a vicious circle of negative learning.

Here's a proposition: *Any business, government, or NGO that depends on dumb scale-up is unlikely to be successful*. Nuclear is just a particularly clear example of this. To secure a future for nuclear, the industry would have to break the current vicious circle of negative learning and crack the code of smart scale-up, with its modularity, replicability, and positive learning-by-doing. Small modular reactors (SMRs) aim to do just that. However, SMRs still lack prototypes and proofs of concept that would allow conclusions regarding whether they are likely to succeed where Big Nuclear has failed. Should small nuclear reactors succeed in this manner, I suggest their name be changed to "*smart* modular reactors."[4]

In the meantime, nuclear's closest competition – renewables – is getting better and better at what they do, resulting in lower electricity prices, which further undermine the business case for nuclear and other Big Energy options. A main reason for the success of renewables is that they have proven well suited for smart scale-up. Wind turbines and solar panels are inherently modular and replicable, and are thus ideal candidates for positive learning and smart scale-up. The first wind turbines were constructed on site, but the nascent industry quickly learned that this was inefficient and changed outdoor construction to indoor manufacturing, based on industrial processes and logistics that could be effectively controlled and rationalized. Today, to talk of the *construction site* for a wind or solar farm, like you talk about the construction site for a nuclear power plant or a big hydroelectric dam, is a misnomer. *Assembly site* better captures what happens when a farm goes up: there is no real construction, just bringing hundreds of wind turbines or thousands of solar panels on site from the factories where they are produced, setting them up one like the other, and connecting them to the grid. This is smart scale-up and it is what makes renewables increasingly competitive, year in and year out.

Here's another proposition: *Any industry, business, government, or NGO that depends on construction sites to deliver their product is unlikely to be successful*. They would have to transform the construction site to an assembly site, with offsite manufacturing of the assembled parts, before they could hope to become truly successful.

---

[4] Thank you to Keivan Samani for the idea to repurpose "SMR" in this manner.

> **Google Fiber: From Smart to Dumb, and Back**
>
> In 2012, Google Fiber was established to provide superfast broadband Internet and TV to US cities. At first, the company tried to scale broadband the conventional way, i.e., by digging cables into the ground. But digging is not one of Google's core competencies, and digging does not scale well, something Google had overlooked. "Digging up streets is definitely not Google's thing," as one analyst dryly observed when it became clear that Google Fiber had hit negative learning and was not delivering as expected. You never know what is underground, which tends to make each dig bespoke and slow, hampering scale-up. In 2016, Google Fiber realized this, stopped digging, and began looking for alternative, more scalable ways to deliver broadband, including wireless. Wireless moves electrons, something Google knows how to do, not earth. Wireless can be delivered in a modular, fast way with standard technology, which is conducive to scaling. Wireless may, in this manner, save Google Fiber from its hubris with bespoke digging and launch the company back into the immaterial world of electrons, modularity, and speed where they need to be to scale successfully (Giwargis and Baron 2016, Wakabayashi 2016).

Renewables have been heavily subsidized, to be sure, but in 2017 a watershed was reached when DONG (later Ørsted), the world's largest offshore windfarm company, decided to build two German windfarms without subsidies, placing further downward pressure on electricity prices and threatening the viability of Big Energy projects like the Hinkley Point C nuclear power station mentioned above (Clark 2017).

Smart scale-up has positive learning, with cost coming down over time. Dumb scale-up has negative learning, with cost going up, like we saw for nuclear power above. In a competition between smart and dumb, as currently witnessed in power production, time is therefore on the side of smart, which means that smart will win and dumb will lose, other things being equal. To the extent that nuclear is going to compete over electricity customers in the future, it looks like it would have to be a competition of small modular reactors against renewables, that is, smart against smart.

**More Examples of Dumb**

Nuclear power is an obvious example of dumb scale-up, almost embarrassingly so. This is because both the technology and the economics of nuclear are so evidently flawed that the deficiencies are impossible to overlook. Other examples of dumb scale-up are less conspicuous in the sense that the engineers succeeded in making the technology work, if with difficulty, and the bad economics have been swept under the carpet or simply forgotten about. The Channel tunnel, the longest under-water rail tunnel in Europe, is an example of this.

The decision to build the privately financed tunnel was made in February 1986, with full passenger service starting almost nine years later, in December 1994. At the initial public offering, Eurotunnel, the private owner of the tunnel, tempted investors by telling them that 10 percent "would be a reasonable allowance for the possible impact of unforeseen circumstances on construction costs" (The Economist 1989: 37-38). After construction began, however, negative learning quickly set in, as is typical for dumb scale-up. The bespoke, one-off design proved much more difficult and costly to build than estimated, both for the tunnel itself and for the trains that would service it. Costs went 80 percent over budget for construction, in real terms, and 140 percent for financing. While costs had to be covered and debt had to be serviced during construction, revenues were years in the future and when they finally materialized, in 1995, they were a fifth of that estimated and never ramped up as projected, even after decades of service. Unforeseen low-cost airlines had undermined the passenger market in unpredictable ways. The laws of forecasting, again. What was planned as a global showcase for private financing of infrastructure quickly ended up a debt-trap, where revenues would never be adequate to service interest payments, not to speak of bringing down debt.

Consequently, Eurotunnel went insolvent and lost billions of investors' dollars, setting back private finance permanently, as a model for infrastructure delivery. The loss to the British economy alone has been estimated at 17.8 billion US dollars, with a negative rate of return on the venture of minus 14.5 percent (Anguera 2006). This is difficult to believe when you use the trains, which are fast, convenient, and competitive with alternative modes of travel. But in fact each passenger is heavily subsidized. Not by taxpayers, this time, but by the thousands of private investors who lost their money during the decade of financial restructuring that was necessary to give the venture a semblance of viability. So next time you use this excellent service, be sure to wear a big smile. Enjoying the ride is the least you can do to warrant the sacrifice of the initial investors and the subsidy they paid you.

The Channel tunnel drives home an important point about dumb scale-up: A large-scale venture done by dumb scale-up – one-off, bespoke, slow – may well be a technological success but an economic disaster, and many are. An ex-post evaluation of the Channel tunnel, which systematically compared actual with forecasted costs and benefits, concluded that "the British Economy would have been better off had the Tunnel never been constructed" (Anguera 2006: 291). The verdict is clear: this is not the kind of venture business or government should be investing in. Why not? Because in addition to wiping out the fortunes of private investors, as described above, it detracts from the national economy instead of adding to it, even when including wider benefits (Vickerman 2017). Rich and big economies like those of the UK and France may be able to absorb large misinvestments like this, because so many other investments are productive. But for smaller and more fragile economies – the majority of nations in the world – just a few bad investments of the size of the Channel tunnel could drag the national economy into the red. Even for large economies, like China, if enough big ventures are done by dumb scale-up – and especially if they are debt financed – they may hurt the national economy (Ansar et al. 2014, 2016).

If you think a case like the Channel tunnel is out of the ordinary, think again. One might have expected that when the Danish Great Belt tunnel – the second-longest underwater rail tunnel in Europe, opened a few years after the Channel tunnel – its owners would have learned their lessons from the latter. But no. At 120 percent in real terms, the Great Belt tunnel had an even larger cost overrun than the Channel tunnel, took longer to build despite being much shorter, and went insolvent before ever carrying a passenger (Flyvbjerg et al. 2003: 13 ff.). Similar examples are depressingly easy to find, from Sydney's Lane Cove and Cross City tunnels over Stockholm's and Oslo's airport express trains to Copenhagen's metro, which all went insolvent by taking a leaf from dumb scale-up's book of long delivery schedules, cost overruns, and benefit shortfalls. Berlin's Brandenburg international airport, opened in October 2020, is on track for similar dismal results, with the mayor forced to step down over the ensuing scandal (The Economist 2014: 29). Again, such assets may be great for those who use them, which makes many think they are great assets as such, when in fact they destroy wealth.

Underperformance like that described applies not only to infrastructure, but to other types of venture as well. For the bespoke Airbus A380 superjumbo, delays, cost overruns, and revenue shortfalls were so bad that the parent company lost more than a quarter of its value and the CEO and other top managers lost their jobs. After more than a decade of sales, the A380 still had not paid for itself and the company realized it never would. In 2019, the company announced production would end by 2021. Boeing and Bombardier have had similar problems of viability with their Dreamliner 787 and C Series, respectively, worsened by the 2020 covid-19 pandemic (The Economist 2015: 59, 2016a: 60). At Kmart, a large US retailer, the entire company went bankrupt when a new, highly customized multi-billion-dollar IT enterprise system, which was supposed to make Kmart competitive with Walmart and Target, went off the rails. The new IT system drained Kmart's finances to a degree where the bankruptcy judge decided to bar the company from making new software acquisitions and put on hold all on-going implementation of the system while the bankruptcy court decided what to do with the company (Flyvbjerg and Budzier 2011).

The failure of the Channel tunnel and the Airbus A380 underscores an issue, which applies to both dumb and smart scale-up: if demand for your product is insufficient, as with these ventures – your product will not succeed. It still matters how you scale, nevertheless. For smart scale-up – with its fast iterations – you learn early whether your product is in demand, and you can cut your losses if it is not. This is what happened with Google Glass, an optical head-mounted camera, computer, and display designed like a pair of eyeglasses and developed by Google X, the research-and-development arm of Google. This seemed a nifty product to geeky Googlers, but a prototype was met with widespread criticism and legal action over privacy and safety concerns, damaging demand and reputation. The backlash was so strong that after only eight months in the market, Google decided to discontinue the prototype. Google undoubtedly lost money as a result, but nothing that would threaten the viability of the company, or even make a dent in its share price, because problems were detected early, before Google Glass had been scaled up. For dumb scale-up, in contrast, fast iterations are not an option and you bet the farm on whatever big, bespoke product it is you are developing. Whether demand will finally support the product will not become clear until the product is finished and fully scaled. If demand is insufficient the consequences are dire, with the whole company suffering or even going bankrupt, as in the case of the privately owned Channel tunnel; or taxpayers losing big sums of money, in the case of public-sector projects.

One could rightly argue that some ventures do not immediately lend themselves to modularization and replication, for instance signature architecture, subways, big bridges, tunnels, and dams. What do you do in that case? Is there a way to avoid the pitfalls of dumb scale-up, nevertheless? The answer is yes. You design and manage things to get as far from the lower left-hand corner of Figure 1, dumb scale-up, as you can, and as close as possible to the upper right-hand corner, smart scale-up, in full knowledge that with a venture that does not fully lend itself to modularization and replication you cannot go all the way. But you can typically go further than you think, and inspirational examples exist of how to do this, like the Empire State Building, the Madrid metro under president Manuel Melis, and architect Frank Gehry's BIM-based, award-winning buildings (Flowers 2009; Lindsey 2001; Lynn 2013; Melis 1999, 2003).

In sum, dumb scale-up is one-off, bespoke, and slow and correlates closely with delays, cost overruns, and benefit shortfalls. Any business, government, or NGO that depends on dumb scale-up is unlikely to ultimately be very successful. Nevertheless, dumb scale-up is a common business model, incredible as it may sound. Dumb scale-up should be avoided, where possible, either by introducing elements of smart scale-up, or by outright disruption, where dumb scale-up is replaced by smart scale-up, as is currently happening in electricity production with smart-scaled wind and solar out-competing dumb-scaled Big Energy.

**Forced Scale-Up and Fumbled Scale-Up**

In addition to smart and dumb scale-up, Figure 1 depicts two further types of scale-up: forced and fumbled. *Forced scale-up*, shown in the upper left-hand quadrant, designates the combination of a one-off, bespoke design and high-speed delivery. This will typically result in a product of low quality, because it is difficult to deliver bespoke designs at speed. For ventures with forced scale-up, speed is typically imposed from the outside, for instance by pressure from top management or politicians for monumental prestige projects, or by the project running up against an immovable deadline, as in the case of mega-events.

Hosting the Olympics or the FIFA World Cup are examples of forced scale-up, with their typical combination of bespoke signature architecture – notorious for being late and over budget – and a deadline written in stone that hosts almost always have difficulty meeting. Cost overrun for the Olympic Games is higher than for any other type of megaproject, at 172 percent on average in real terms (Flyvbjerg et al. 2020). Many host countries and cities are littered with low-quality stadiums and other facilities that are underutilized or boarded up. Facilities often begin falling apart immediately after events are over, many ending up as full-fledged white elephants. Athens, Beijing, and Rio de Janeiro are recent examples, as host cities for the Summer Olympics. The Athens 2004 Games, hosted by Greece – a small country with a fragile economy – became a contributing factor to the country's 2011 debt default. The cost overruns and incurred debt from the Games were so large that they negatively affected the credit rating of the whole nation, thus weakening the economy in the years before the 2008 international financial crisis. This resulted in a double-dip recession in Greece – with financial and economic disaster still rippling through the economy a decade later – where other nations had only a single dip and recovered faster.

China's construction bubble – driven by a policy of nation building through accelerated development of infrastructure, buildings, and whole cities – may also be considered an instance of forced scale-up (Ren 2017). To illustrate just how accelerated the speed is, consider that China poured more concrete in the three years from 2011 to 2013 than the US did the entire $20^{th}$ century. Or that in the five years from 2004 to 2008 China spent more on infrastructure in real terms than it did in the whole of the previous century; that is an increase in the spending rate of a factor twenty. Or, finally, that from 2005 to 2008, China built as many miles of high-speed rail as Europe did in two decades, when Europe was especially busy building this type of infrastructure. At the time of writing, China had 22,000 miles (35,000 kilometers) of high-speed rail lines, more than the rest of the world combined. For decades, China has been driving what *The Economist* calls "the biggest investment boom in history," recently accelerated by President Xi's so-called Belt and Road Initiative, exporting China's model for infra-

                                                                                                                                                                                                                                                                          

structure building on a global scale (The Economist 2008: 80; 2017: 53). Global infrastructure spending has never been this high before in both absolute and relative terms, i.e., measured as a share of world GDP.

Again, this has led to low-quality products and underutilized capacity, such as high-speed rail lines that run trains at lower-than-planned speeds, new roads with little traffic, and whole ghost cities standing vacant with scant prospects of ever delivering a return on capital. And again, if you scratch the proud façade of turbo engineering that China likes to present to the world, real human drama emerges with people losing their jobs and lives and whole government institutions collapsing. For example, in January 2014 Bai Zhongren, president of China Railway Group, jumped to his death in a seeming suicide linked to mismanagement of China's largest and most prominent megaproject, the $329 billion nationwide high-speed rail system. Three years before, in July 2011, two high-speed trains collided at Wenzhou, killing 40 people. Investigators placed responsibility on the former railway minister, Liu Zhijun, who had led the effort to build nearly 5,000 miles (8,000 kilometers) of high-speed rail in seven years, China's most ambitious construction initiative ever. In March 2013, it was decided that the giant Ministry of Railways, with more than two million employees, would be dissolved, and in July Liu was given a suspended death sentence for abuse of power and taking bribes. Ma Cheng, the former head of China Railway Signal and Communication Corporation, was also singled out as responsible for the Wenzhou accident. But Ma was not present to stand trial. He had died of a heart attack a month after the collision. Following the indictment of Liu, Ma, and dozens of other rail officials, Bai Zhongren was up next for investigation. He apparently caved to the pressure and jumped to his death (LaFraniere 2011, BBC 2013, The Economist 2013, Wang 2014).

Dumb scale-up is bad. Forced scale-up is worse. This is because the combination of speed and low quality wastes resources faster and at a larger scale than dumb scale-up. Forced scale-up therefore wrecks even more havoc than dumb scale-up, as illustrated by the examples above. In September 2014, *The South China Morning Post* reported that even China, which for decades had emphasized a policy of "speed-at-all-costs" was beginning to see the downside of forced scale-up and was now "more willing to accept delays in building equipment and infrastructure to ensure quality" (Chen 2014).

Finally, *fumbled scale-up* is shown in the lower right-hand quadrant of Figure 1. This is the least bad of the three types of non-smart scale-up. Fumbled scale-up designates a combination of modular, replicable design and low speed in delivery. In principle, modularity would allow for high-speed rollout, but for whatever reason speed is not achieved, for instance because of lack of capital, person-nel, C-suite attention, or other resources. The result is a missed opportunity. If competitors exist that master smart scale-up, they will typically outcompete the entity that fumbled their scale-up.

As an example of fumbled scale-up, consider the joint venture of UK supermarket chain Sainsbury's and Danish low-cost retailer Netto to roll out discount stores in the UK in the mid 2010s. Discount stores are modular and replicable and may be built at speed. Nevertheless, after three years Sainsbury's and Netto had established only 16 stores in the UK, while the competition – Germany-based Aldi and Lidl – had set up hundreds with hundreds more planned. What happened? Sainsbury's got sidetracked by a major acquisition and lost focus. Moreover, availability and cost of appropriate sites for discount stores caused difficulties, because of increased competition and demand. As a consequence, Sainsbury's and Netto were slowed down, which is the one thing you cannot afford in a competitive market where smart scale-up decides who succeeds.

Mike Coupe, British CEO of Sainsbury's, rightly concluded, "To be successful over the long term, Netto would need to grow at pace and scale" (*The Guardian* 2016). Per Bank, Danish CEO of Netto, similarly observed, "the business requires greater scale over a short period of time to achieve long-term success." Both CEOs knew that pace and scale were the solution to their problems, but pace and scale were not forthcoming so they had no other choice but to leave the market to their competitors, close down the joint venture, lay off several hundred people, and write off their investment, which is what they did. Sainsbury's lost the opportunity to enter discount retailing and get a more diversified product portfolio. Netto lost a much coveted opportunity to enter the UK market after success in several other nations. Both businesses wasted substantial time and money because they fumbled the scale-up process. Scale-up takes no prisoners when competition is fierce, as it is in discount retailing. If you fumble you die.

**From Dumb to Smart Scale-Up**

You can make yourself and your organization – whether a business, a government agency, or an NGO – hugely more valuable if you, first, become clear about which of the four quadrants in Figure 1 you currently operate in and, second, move your focus and your activities systematically and effectively towards the upper right-hand side, to smart scale-up. This is not an either-or proposition but a matter of degree. Your organization and most of its ventures – existing or planned – will be neither fully dumb nor fully smart, one or the other, but will have elements of each. Your task is to increasingly and tenaciously tip the balance towards smart by beginning to introduce smart-scaled ventures, and elements of smart scale-up into existing ventures, to make your organization less dumb and ever smarter.

> **Ørsted: Pivot from Dumb to Smart**
>
> In 2017, DONG, the world's largest offshore windfarm company, decided to build two German wind farms without subsidies. This sent shock waves through the global energy industry. It marked the first time in history that offshore wind became cheaper than fossil fuels, placing disruptive pressure on electricity prices and threatening the viability of conventional energy sources. The low prices were made possible by two decades of relentless efforts by the wind industry to standardize, modularize, and enlarge wind-turbines to a degree where multi-billion-dollar mega farms could be installed in a matter of months, like giant sets of LEGOs, driving down the cost of producing electricity to levels never seen before. In comparison, multi-billion-dollar coal, gas, hydro, and nuclear power plants are widely bespoke and take years to build, leaving costs non-competitive and Big Energy open to disruption. Tellingly, shortly after winning the two German wind farms DONG decided to pivot and sell its billion-dollar oil and gas business, from which it was named (DONG is short for Danish Oil and Natural Gas), and to rechristen itself Ørsted – after Danish physicist Hans Christian Ørsted, who discovered electromagnetism – with a new company mission "to create a world that runs entirely on green energy" (Clark 2017, Spector 2017).

Specifically, you need to ask, "How can we make each thing we do more modular and less bespoke, more replicable and less one-off, more speedy and less slow?" or, in short, as expressed in the following fast-and-frugal heuristic:

*Ask yourself, how can we do things more modular and faster, now?*

If you do this – and if you get your team and wider organization to adopt this approach – you will unleash potential you never knew was there. You will be able to change your business and impact the world in ways you did not think possible, like the 20,000 schools did for hundreds of thousands of children in Nepal and like Tesla is doing for myriad consumers who vote with their feet for new lifestyles in energy and transportation. That's smart!